\documentclass[sigconf]{acmart}
\usepackage{graphicx} 
\newcommand{\cmss}[1]{{\fontfamily{cmss}\fontseries{c}\selectfont{#1}}}

\AtBeginDocument{%
  }

\copyrightyear{}
\acmYear{}
\setcopyright{none}
\setcctype{by}
\acmConference{}{}{}
\acmBooktitle{}
\acmPrice{}
\acmDOI{}
\acmISBN{}

\renewcommand\footnotetextcopyrightpermission[1]{} 

\begin{document}
\title{Graphical-Probabilistic Modeling of Generative Flows in LLM-Native Software Systems}

\author{Víctor A. Braberman}
\email{vbraber@dc.uba.ar}
\orcid{0000-0001-5946-3550}
\author{Flavia Bonomo-Braberman}
\orcid{0000-0002-9872-7528}
\email{fbonomo@dc.uba.ar}
\affiliation{%
  \institution{Departamento de Computación, FCEN, Universidad de Buenos Aires / ICC, UBA-CONICET}
  \city{Buenos Aires}
  \country{Argentina}
}

\date{October 2025}
\begin{abstract}
    Engineering LLM-native software remains a challenging and immature field. Current practice is largely exploratory, relying on experimentation and heuristic techniques such as prompting and context engineering. These, however, are low-level and lack the principled structure needed to support design-level reasoning or analysis.
In contrast, traditional software engineering leverages modularity and abstraction to communicate and analyze system behavior. To bring similar rigor to LLM-native development, we propose methods for documenting generative flows and for stating properties of LLM-based software designs. 
Such methods must account for the stochastic, prompt-dependent behavior of large language models while remaining expressive enough to capture emergent phenomena.
Our initial approach is based on graphical probabilistic models, tailored to capture phenomena characteristic of LLM-native systems. This framework—what we term Generation Networks—aims to provide a foundation for principled reasoning about generative interactions and system-level properties in LLM-centric software architectures.

\end{abstract}

\ccsdesc[500]{Software and its engineering~System description languages}
\ccsdesc[500]{Software and its engineering~Software verification and validation}
\ccsdesc[500]{Computing methodologies~Probabilistic reasoning}
\ccsdesc[500]{Software and its engineering~Extra-functional properties}

\keywords{AI-Enabled Systems, LLMs, LLM-native applications, design specification, graphical probabilistic models, Generative Networks}

\maketitle
\section{Introduction}
Engineering software systems that incorporate large language models (LLMs) remains a difficult and largely ad-hoc activity~\cite{DBLP:journals/pacmse/LiangLRM25}. Development practices rely on heuristics—prompt engineering, context tuning, and workflow scripting—that work in isolation but provide little support for system-level reasoning or design documentation. The absence of abstraction mechanisms comparable to those in traditional software engineering hampers analysis, maintenance, and reuse.

Recent conceptual work has been proposed viewing LLM-enabled applications—systems where LLMs perform essential computational roles—as collections of LLM-mechanized,  potentially interleaved with external tool calls, orchestrated
through arbitrary control flow (e.g., transformations~\cite{braberman2024tasks,DSPy,DBLP:journals/corr/abs-2507-19457}). 

This paper extends that line of thought 
showing how probabilistic modeling can describe  generative software designs. More concretely, we introduce  graphical models tailored to LLM-based systems, to document data flows and express properties of interest. The examples highlight how conceptual variables, distributional parameters, and dependency structures can encode notions of correctness, robustness, and design improvement.

Our aim is to show how these modeling constructs provide an engineering vocabulary for articulating the structure and behavior of LLM-enabled systems.
By integrating software-engineering abstractions with the formalism of probabilistic modeling~\cite{DBLP:books/daglib/0023091}, the proposed approach—termed Generation Networks—offers a principled framework for communicating, documenting, and reasoning about architectures whose behavior emerges from stochastic, prompt-sensitive generative mechanisms.

The paper is structured as follows. Section 2 reviews related work, and Section 3 shows the basics of Generative Networks and illustrates its use in documentation. Section 4 introduces a language for specifying design properties. Section 5 presents use cases that combine these constructs. Finally, Section 6 outlines future research directions, and Section 7 concludes by discussing the implications for LLM-native software development.

\section{Related Work}

Graphical probabilistic models~\cite{DBLP:books/daglib/0023091} have been argued to represent neural inference at different levels of abstractions and for different communication goals. 
\paragraph{Hypothesis regarding internal mechanisms:} Causal models, causal abstraction, and causal graphs has been applied to interpretability of neural networks activation~\cite{DBLP:conf/nips/GeigerLIP21}. There mechanisms underlying network inference for one requested task are hypothesized by using causal graphs.
\paragraph{Hypothesis regarding prompt-code relationship:} In~\cite{DBLP:journals/pacmse/JiMLWW25} it is proposed a causal graph-based representation of the prompt and the generated code, which is established over the fine-grained, human-understandable concepts in the input prompts used specifically for code generation.
\paragraph{Uniform blueprint of existing  techniques:} In~\cite{cascades} authors argue that existing LLM techniques like scratchpads~\cite{nye2021workscratchpadsintermediatecomputation}, CoT~\cite{CoT-Wei22}, verifiers~\cite{cobbe2021trainingverifierssolvemath}, STaR~\cite{DBLP:conf/nips/ZelikmanWMG22}, selection-inference and tool use~\cite{DBLP:conf/iclr/CreswellSH23}, can be expressed as particular probabilistic programs (cascades), and therefore represented in the language of graphical models with random variables whose values are complex data types such as strings.
\paragraph{Uniform blueprint of LLM-native Software Designs:}

In~\cite{braberman2024tasks}, a systematic analysis of over one hundred LLM-based solutions from the software engineering literature was conducted to characterize the abstract functionalities assigned to LLMs. This was achieved by analyzing concrete prompt-based interactions to recover the underlying intent of each task (a.k.a, transformation). 
While that study primarily focused on identifying a typology of transformations and recurring composition motifs, it proposed—as a future research direction—the unification of LLM-native system descriptions through data-dependency graphs. Under this abstraction, the proposed ``world model'' provides a schematic representation of the system's behavior in terms of its input, internal, and generated data variables. This perspective serves as the core inspiration for the approach adopted in this paper to represent and analyze LLM-integrated architectures.

\begin{figure}[t]
\centering
\includegraphics[width=.98\columnwidth]{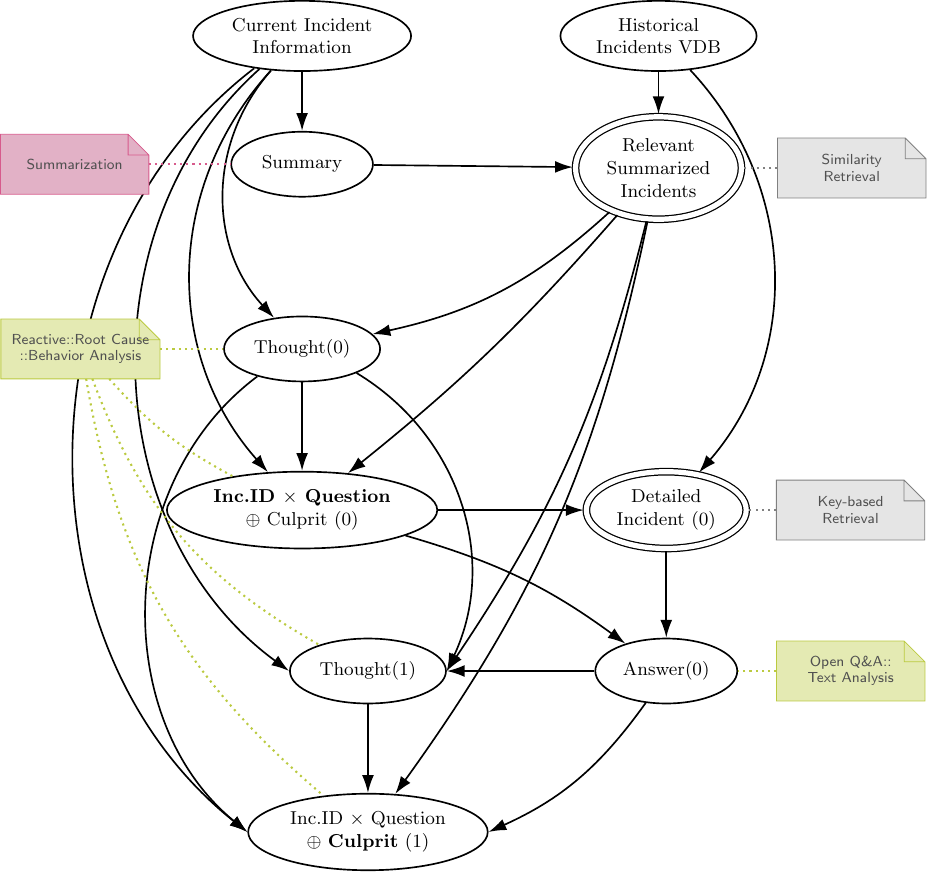}
\Description{A computational graph for an incident response system with 10 variables and 5 process notes. 
Variables (represented as nodes): (1) "Current Incident Information"; (2) "Summary"; (3) "List of Relevant Summarized Incidents"; (4) "Historical Incidents DB"; (5) "Thought(0)"; (6) "(Inc.ID x Question)" xor symbol "Culprit" (7) "Answer(0)"; (8) "Thought(1)"; (9) "(Inc.ID x Question)" xor symbol "Culprit", (10) "Detailed Incident (0)". Node characteristics: (3) and (10) are double circled since they are deterministic variables. In (6) boldface "(Inc.ID x Question)" part of the label while in (9) boldface "Culprit".
Edges: (1)->(2), (4)->(3), (2)->(3), (1)->(5), (3)->(5), (5)->(6), (6)->(10), (4)->(10), (10)->(7), (6)->(7), (7)->(8), (1)->(8), (3)->(8), (8)->(9), (1)-> (6), (3)->(6), (1)->(9), (3)->(9), (7)->(9), (5)->(8), (5)->(9). 
Note nodes (a) "Summarization" linked with dotted line as in reference with (2); (b) "Similarity Retrieval" linked to (3); (c) "Key-based Retrieval" linked to 10; (d) "Reactive::Root Cause::Behavior Analysis" linked to (5), (6), (8), (9); (e) "Open Q&A::Text Analysis" linked to (7)}
\caption{
An abstract execution of a hypothetical RCA tool
modeled using transformation types as categorized by~\cite{braberman2024tasks}. }
\label{fig:RCA}
\end{figure}

\section{Design Representation via Data-Dependency Graphs}

A primary goal of this proposal is to communicate the central design decisions of a system—whether currently operational or under specification—by modeling some representative executions as Directed Acyclic Data-Dependency Graphs (DDGs). In these graphs, nodes denote conceptually typed data instances—comprising input, output, and internal variables—while edges reify the data transformations~\cite{braberman2024tasks} that connect them. 
A transformation  maps a set of parent variables values to a target child variable value. In LLM-native software, these mechanisms can be categorized into two primary types:
\begin{enumerate}
    \item \textbf{LLM-based transformations}: These stochastic transformations abstract prompt-modulated interaction where the LLM produces the content of the target variable. Those prompts typically would include instructions, examples, and string-like representations of the parent variables values. The conditional probability distributions within these transformations (prompt continuations) are ultimately defined implicitly by the opaque process of token generation\footnote{This is the emerging result of the specific network architecture~\cite{vaswani2017attention}, its fixed weights, the effects of chosen generative/sampling adapters, and hyperparameters~\cite{10.1162/tacl_a_00502}.}.
    
    \item \textbf{Algorithmic transformations}: Typically deterministic processes, such as data projection, composition, or similarity-based retrieval from vector databases.
\end{enumerate}

\paragraph{Example.} \autoref{fig:RCA} illustrates an abstract execution scenario of a hypothetical LLM-based Root Cause Analysis (RCA) tool inspired by~\cite{RCACopilot}. This tool serves as a representative Retrieval-Augmented Generation (RAG)~\cite{RAG2020Patrick} agent designed for IT Incident Response. The system first summarizes the current report and log, retrieves similar historical cases via algorithmic similarity-search to establish context, and then iteratively decides whether to query a specific historical incident or identify the ``Culprit'' by following a ReAct-style reasoning–action loop~\cite{ReAct} with tool-mediated actions. The representative execution depicted here is an execution that reaches a culprit after requesting and checking further details regarding what the LLM considers to be the closest registered incident. 
In this scenario, \texttt{Current Incident Information} serves as an input variable, while nodes such as \texttt{Summary}, \texttt{Thought(i)}, and \texttt{Answer(i)} represent variables produced through transformations\footnote{As execution involves a trajectory of values for the same conceptual variable, names are postfixed with an instance number to identify each realization within the trace.}. Generative transformations—labeled according to types such as \cmss{Summarization} or \cmss{Behavior Analysis}~\cite{braberman2024tasks}—yield stochastic content through LLM-mediated processes. Conversely, algorithmic transformations (e.g., similarity retrieval) produce deterministic content, denoted by double-circled nodes such as \texttt{Detailed Incident~(0)} and \texttt{Relevant Summarized Incidents}. Consider the variable \texttt{(Inc.ID $\times$ Question) $\oplus$ Culprit~(1)}. Its realization depends on the initial incident report, the retrieved summaries, and the history of generated thoughts and obtained answers. Its distribution is governed by the \cmss{Behavior Analysis} transformation, which abstracts a prompt-based interaction instantiated with the values of these parent variables. Crucially, as an LLM-based transformation, it involves the autoregressive generation of \texttt{Thought(1)}—the verbalized Chain of Thought~\cite{CoT-Wei22}—which also  conditions the subsequent probability of the utterances for the final action. Thus, the network clarifies how the system leverages iteratively generative capabilities to reach a diagnostic conclusion, defining the precise generative contexts even without exhibiting concrete prompting details.

\paragraph{From DDGs to BNs.}The mapping from a DDG to a formal probabilistic model is firstly grounded in the nature of system execution. Given the intrinsic stochasticity of input and LLM-based transformations, each node in the DDG—comprising input, internal, and output data—can be formally regarded as a random variable. Moreover, for any fixed DDG, there exists a set of execution traces (i.e., sets of random variables values) consistent with its specific topology: the graph structure itself often encodes termination criteria or specific logical paths (e.g., a bounded unfolding of an iterative structure that actually concludes when culprit is yielded). 

This allows for the derivation of a corresponding Bayesian Network (BN)~\cite{DBLP:books/daglib/0023091} to represent the distribution's conditional independence structure for data compatible with the given DDG. In this framework, the variable dependencies define the causal relations~\cite{Pearl_2009} among variables. 

\section{Modeling Design Properties}

In addition to conveying structural task decomposition, we extend the representational and denotational scope of our notation, Generation Networks, to capture more nuanced design-level properties. This expansion is specifically tailored to support the rigorous documentation and analysis of LLM-based software architectures, moving beyond simple workflow visualization to a formal modeling of the interaction between deterministic logic and stochastic generation.
\paragraph{Distribution-control and domain selector variables.}
In previous examples, details such as the specific model architecture, weights, or prompt instructions were left implicit within the underlying conditional distribution governing stochastic transformations.
In some situations, however, it is useful to expose these factors explicitly—e.g., (typically latent) model parameters~$\theta$ or controllable variables such as temperature, hyperparameters, or fixed demonstrations and instructions.

Likewise, it can be convenient to introduce \emph{domain selector} variables that modulate the distribution of exogenous variables, capturing which acquisition or data-source regime is being considered (e.g., benchmark, synthetic, or in-the-wild~\cite{Glocker20}).  
Such variables enable reasoning about \emph{acquisition shifts}: situations where transformations must operate on data drawn from different representational or contextual distributions while addressing the same conceptual problem.  
For instance, when modeling a defect-detection transformation, one may distinguish between abstract categories of a defect (e.g., a class of memory-bound violations) and their concrete syntactic realizations across programming languages or styles.  
This distinction allows designers to articulate expected behavior under distributional variation, analyze robustness to representational changes, and document assumptions about training versus deployment conditions.

Accordingly, we extend the notation with elements of meta-network modeling~\cite{DBLP:books/daglib/0023091}, allowing selected distribution-control variables—whether affecting exogenous or generated variables—to be represented explicitly when relevant.  
This representation clarifies the use cases introduced later and generalizes naturally to modules implemented by other neural models beyond LLMs.

\paragraph{On the Modeled World.}
Complementing the representational extensions above, we further expand the denotational scope of the framework.  
Beyond variables that denote observable inputs or data generated during execution, Generation Networks allow the inclusion of conceptual or latent quantities required to express asserted or expected system properties.  
For example, a diagram may introduce variables representing the (typically unobservable) ground-truth or aligned outputs of transformations, as well as virtual intermediate results—such as Boolean indicators capturing whether an outcome aligns with its expected value.  
These variables enable the specification of evaluative properties, including expected success rates, compositionality relations, and robustness conditions under given assumptions.

\paragraph{Specifying Quantities}
A key extension of Generation Networks is the ability to express \emph{first-order formulas} defined over functional symbols that denote quantities derived from the joint probability distribution of the modeled variables. 
These formulas build on two classical operations from causal modeling~\cite{DBLP:books/daglib/0023091}: 
\emph{probabilistic queries} and \emph{intervention (causal) queries}. 
Through them, Generation Networks can state relations between probabilistic quantities, enabling the formal expression of prescriptions, assertions, and search conditions based on the system’s joint distribution.

A \emph{probabilistic query} represents the posterior probability that variables \(Y\) take specific values given evidence \(E=e\):
\[
P_N(Y \mid E = e)
\]
An \emph{intervention query} expresses the posterior distribution after enforcing a manipulation on certain variables, capturing causal reasoning in Pearl’s sense:
\[
P_{N_{Z=z}}(Y \mid E = x)
\]

Thus, together, these specification-level constructs  express design desiderata:  descriptive, predictive, and prescriptive statements about LLM-native system designs modeled through Generation Networks. 
As said, these functionals denoting distributions can be used to express point-wise comparisons or divergence comparisons (e.g., KL~\cite{kullback1951information} or Wasserstein Distance~\cite{DBLP:conf/icml/ArjovskyCB17}) between those probabilistic quantities.

\begin{figure}
    \centering
    \includegraphics[width=0.9\linewidth]{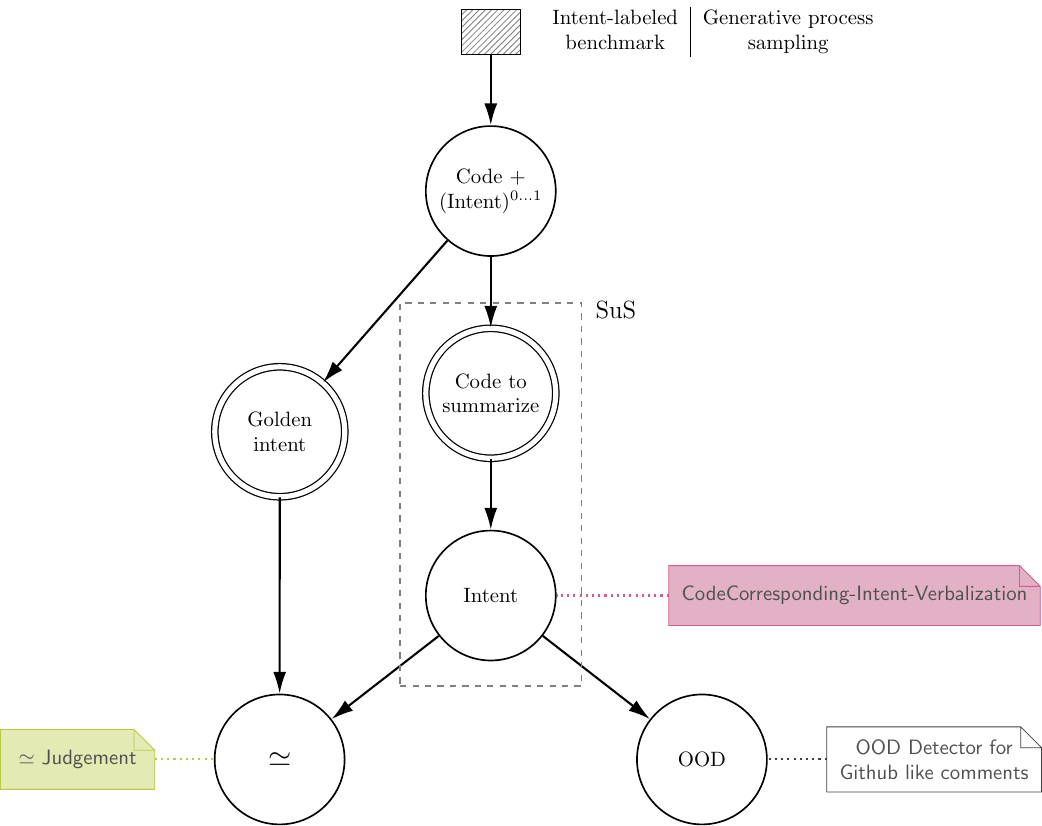}
    \caption{Prescription Use Case}
    \label{fig:prescribing}
    \Description{A vertical computational data-dependency graph organized by a central processing spine and a defined system boundary.

1. \textbf{Source and Input:} At the top center, a shaded gray rectangle represents the ``Intent-labeled benchmark,'' serving as the entry point for generative process sampling.

2. \textbf{The Central Processing Spine:} A vertical sequence of solid arrows connects three nodes. The top node is a single-ellipse (stochastic) node labeled ``Code + Intent.'' This points down to a double-circled (deterministic) node labeled ``Code to summarize,'' which in turn flows into a single-ellipse node labeled ``Intent.''

3. \textbf{The System under Specification (SuS):} A dashed rectangular box labeled ``SuS'' encloses the ``Code to summarize'' and ``Intent'' nodes. This boundary distinguishes the internal system components from the external evaluation framework.

4. \textbf{Evaluation and Comparison Logic:} 
   \begin{itemize}
       \item To the left of the spine, a double-circled node labeled ``Golden intent'' receives an arrow from the top ``Code + Intent'' node.
       \item At the bottom-left, a node labeled with a tilde-equals symbol ($\simeq$) represents the ``Judgement'' variable. It receives converging arrows from both the ``Golden intent'' node and the system's generated ``Intent'' node.
       \item At the bottom-right, an ``OOD'' (Out-of-Distribution) node receives an arrow from the system's ``Intent'' node.
   \end{itemize}

5. \textbf{Functional Annotations:} Three document-shaped notes are linked via dotted lines to specify the underlying transformations:
   \begin{itemize}
       \item A pink note for ``Code-Corresponding Intent Verbalization'' is linked to the ``Intent'' node.
       \item A green note for ``Judgement'' is linked to the comparison node ($\simeq$).
       \item A gray note for ``OOD Detector'' is linked to the ``OOD'' node.
   \end{itemize}
}

\end{figure}
\section{Use Cases}

We now explore some potential uses of the modeling framework (some of them speculative)  by combining the modeling features mentioned above. These use cases schematically illustrate framework's expressive power and how it enables a uniform and principled way to treat a wide variety of design communication/inquiring operations. Due to a lack of space, we choose to illustrate some subset of uses. 
  
\subsection{Prescribing Transformation Behavior}

\autoref{fig:prescribing} illustrates how probabilistic prescriptions can formalize expected behavior of a transformation, in this case, the \cmss{CodeCor\-re\-spond\-ing-Intent-Verbalization} task.
The exogenous variable $\mathit{Code{+}Intent}^{0\ldots1}$ supplies a code snippet paired, optionally, with a reference (“golden”) intent description.
This input is deterministically projected into these two elements. Generated random variables include:  (i)~an $Intent$ variable representing the verbalized intent produced by the transformation under prescription, and (ii)~a Boolean variable $\simeq$ capturing the outcome of an LLM-based correctness judgment (true if the generated intent matches the golden reference).
Additionally, an $OOD$ variable encodes the output of an out-of-distribution detection mechanism (e.g.~\cite{wu-etal-2023-multi,DBLP:conf/iclr/0006LZKSLL23}), indicating whether the generated intent resembles text sampled from an expected domain (e.g., intents from GitHub projects).
A categorical domain selector, $D$, distinguishes between two regimes:
(1)~a benchmark regime, where golden intents are available, and
(2)~a generative regime, where code snippets are sampled from a learned generative process.
For each regime, we specify probabilistic prescriptions reflecting minimal performance criteria:
$$[
P(\simeq = \text{True} \mid D = \text{benchmark}) \ge 1 - \varepsilon
]$$
$$[
P(OOD = \text{True} \mid D = \text{generative}) \le \delta
]$$
The first ensures near-correct behavior on benchmark data; the second enforces that generated intents remain within an acceptable distributional range (i.e., resemble legitimate comment-like text).
These prescriptions exemplify how the proposed representation can encode behavioral expectations as constraints on conditional probabilities over transformation variables.

\begin{figure}
    \centering
    \includegraphics[width=0.9\linewidth]{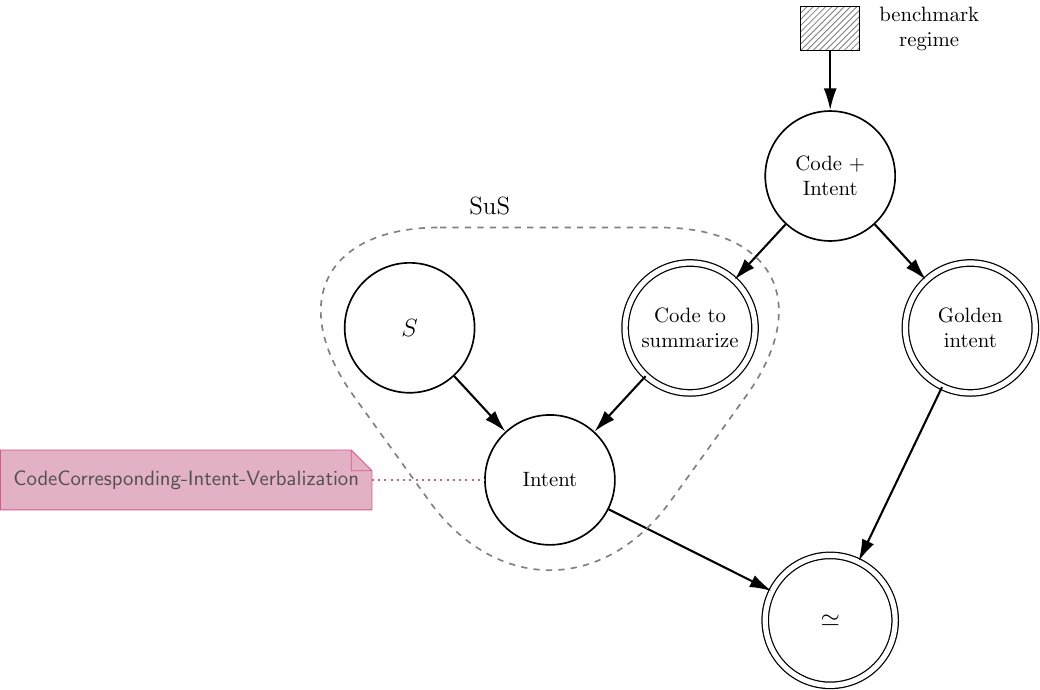}
    \caption{Exploration Use Case}
    \label{fig:exploration}
    \Description{A computational data-dependency graph with a triangular layout, representing an evaluation flow for code summarization.

1. \textbf{Source and Input:} At the top center, a shaded gray rectangle represents the ``benchmark regime.'' It flows into a single-ellipse (stochastic) node labeled ``Code + Intent.''

2. \textbf{Data Branching:} From the ``Code + Intent'' node, the graph splits into two paths:
   \begin{itemize}
       \item To the left, a double-circled (deterministic) node labeled ``Code to summarize.''
       \item To the right, a double-circled (deterministic) node labeled ``Golden intent.''
   \end{itemize}

3. \textbf{The System under Specification (SuS):} A large, dashed triangular boundary labeled ``SuS'' encloses the core system logic. Inside this boundary are:
   \begin{itemize}
       \item The ``Code to summarize'' node.
       \item A single-ellipse node labeled ``$S$,'' representing few-shot examples or context.
       \item A single-ellipse node labeled ``Intent,'' which receives converging arrows from both ``Code to summarize'' and ``$S$.''
   \end{itemize}

4. \textbf{Evaluation and Convergence:} At the bottom center, a double-circled node labeled with a tilde-equals symbol ($\simeq$) represents the final comparison or judgment. It receives converging arrows from the system-generated ``Intent'' (bottom-left) and the ``Golden intent'' reference (top-right).

5. \textbf{Functional Annotation:} A pink document-shaped note labeled ``Code-Corresponding Intent Verbalization'' is linked via a dotted line to the central ``Intent'' node, identifying the transformation mechanism.
}
\end{figure}

\subsection{Requesting the Exploration of Mechanism Parameters}
Having introduced probabilistic prescriptions that express expected behavior of a transformation, we now turn to the complementary question of how such behavior can be modulated through explicit mechanism parameters.
\autoref{fig:exploration} illustrates how parameters influencing the behavior of neural-based transformations can be made explicit within the representational model.
Consider again the \cmss{CodeCorresponding-Intent-Verbalization}  transformation, where a code snippet and its corresponding golden intent are available under a benchmark regime.
We introduce a variable $S$ denoting the set of few-shot examples~\cite{Few-Shot-Brown20} included in the prompt—along with the current code snippet—to guide the model’s generation of the predicted \emph{Intent}. Again, $\simeq$ is a similarity judgment between the result and the golden intent. In this use case, the designer requests optimization with respect to available benchmark data\footnote{This is akin to declarative specification frameworks such as DSPy~\cite{DSPy}.}.  
In our framework, exploring alternative few-shot configurations corresponds to performing interventions on \(S\), yielding post-intervention distributions \(P_{N_{S=s}} (\simeq)\) and the search for an optimal configuration can be formalized as a model-based optimization problem:
\[
s^\star = \arg\max_{s \in \mathcal{S}} \; P_{N_{S=s}}(\simeq = True).
\]

In this form, Generation Networks treat prompt configuration and similar hyperparameters as interventional variables, enabling declarative specification of search, optimization, or tuning tasks within the same probabilistic framework used for descriptive and causal reasoning. In fact,  conceptual prompt evaluation and calibration is another related use case that could be approached within our framework, similarly to what \cite{DBLP:journals/pacmse/JiMLWW25} does for code generation.  
Beyond prompt-level optimization, our formalism supports exploratory \emph{what-if} analyses for design reasoning.  
For example, robustness to systematic component degradation or failures can be investigated by intervening on selected nodes of a design-level network, thus simulating alternative operational scenarios within the same representational language.

\begin{figure}
    \centering
    \includegraphics[width=0.87\linewidth]{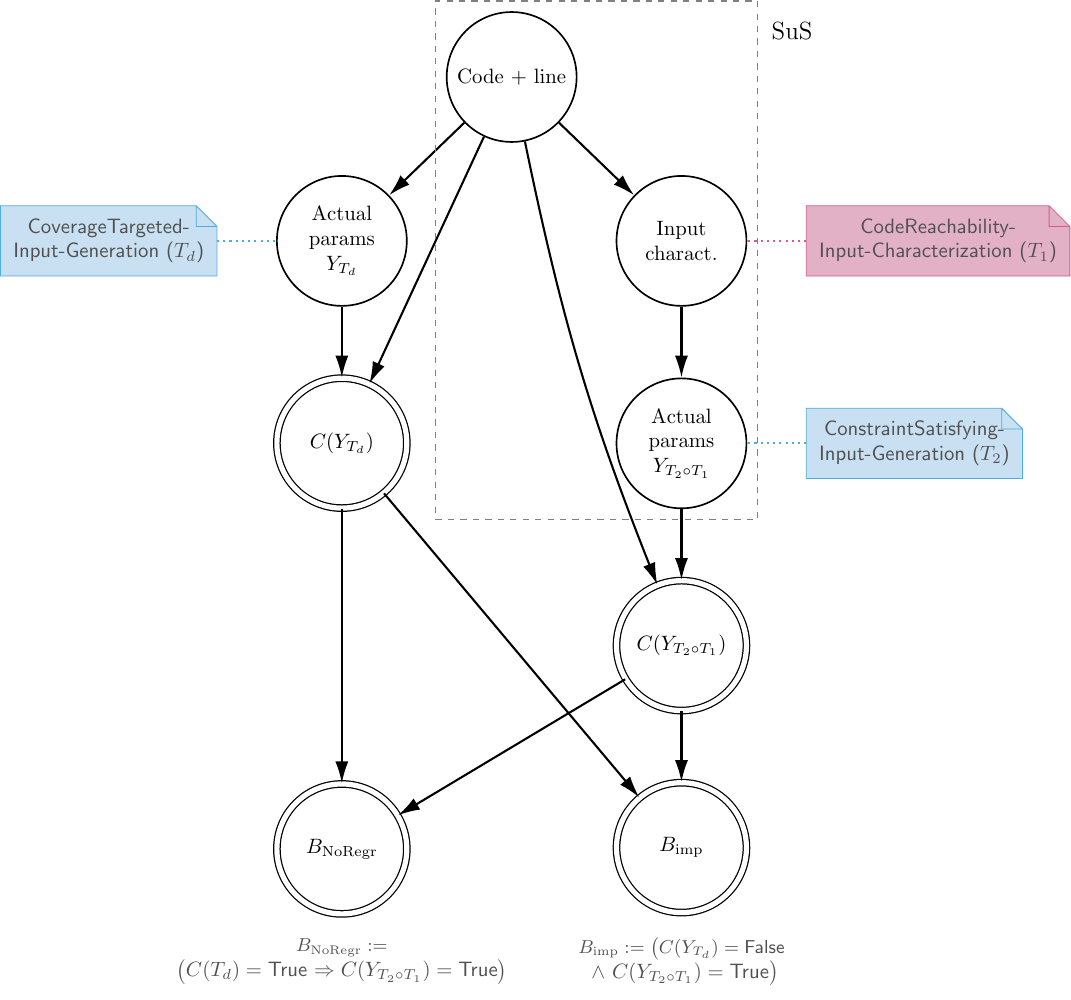}
    \caption{Improvement Use Case}
    \label{fig:improvement}
    \Description{A hierarchical data-dependency graph modeling two parallel strategies for input generation and their evaluation against Boolean predicates.

1. \textbf{Source and System Boundary (SuS):} At the top center, the stochastic node ``Code + line'' serves as the common input. A dashed gray rectangular box labeled ``SuS'' encloses this source node along with the right-side branch: ``Input charact.'' and ``Actual params $Y_{T_2 \circ T_1}$.''

2. \textbf{Left Branch (Direct Generation):}
   \begin{itemize}
       \item The source node points to a stochastic node ``Actual params $Y_{T_d}$.''
       \item Both nodes flow into a double-circled (deterministic) node ``$C(Y_{T_d})$,'' representing a coverage check.
       \item A blue document note ``CoverageTargeted-Input-Generation ($T_d$)'' is linked to the generation node.
   \end{itemize}

3. \textbf{Right Branch (Chained Generation):}
   \begin{itemize}
       \item The source flows into ``Input charact.'' (stochastic), annotated with a pink note ``CodeReachability-Input-Characterization ($T_1$).''
       \item This flows into ``Actual params $Y_{T_2 \circ T_1}$'' (stochastic), annotated with a blue note ``ConstraintSatisfying-Input-Generation ($T_2$).''
       \item These nodes, plus a direct link from the source, converge on a double-circled node ``$C(Y_{T_2 \circ T_1}$),'' representing the coverage result for the chained strategy.
   \end{itemize}

4. \textbf{Evaluation Predicates:} At the bottom, two double-circled deterministic nodes receive arrows from both the direct coverage node and the chained coverage node:
   \begin{itemize}
       \item Node ``$B_{\mathrm{imp}}$'': Represents an Improvement predicate. An unbordered note defines it as the case where direct generation fails coverage but chained generation succeeds.
       \item Node ``$B_{\mathrm{NoRegr}}$'': Represents a No-Regression predicate. An unbordered note defines it as the logical implication that if direct generation succeeds, chained generation must also succeed.
   \end{itemize}
}
\end{figure}

\subsection{Asserting Design Improvement}

Generation Networks can also express design improvements in realistic transformation pipelines through deterministic Boolean variables that depend on observable performance outcomes. ~\autoref{fig:improvement} sketches this use case. Consider a transformation ($T_d$) performing \cmss{CoverageTargeted-Input-Generation}: given a parameterized code snippet and a target line, it attempts—within a single inference—to produce an input that triggers execution of that line.
A decomposed design instead realizes the same goal through two sequential transformations:
($T_1$), \cmss{CodeReachability-Input-Characterization}, which verbalizes a natural-language description of inputs expected to reach the target line; and
($T_2$), \cmss{ConstraintSatisfying-Input-Gen\-er\-a\-tion}, which generates a concrete input consistent with that description.
For both designs, goal achievement is deterministically signaled by checking whether the target line is actually covered during execution—no golden reference is required.

Let $Y_{T_d}$ and $Y_{T_2\circ T_1}$ denote the respective outputs (generated inputs) for the same code–line pair, and let a deterministic variable  $C(\cdot)$ indicate whether the resulting input achieves coverage.
Then a Boolean variable capturing relative correctness between the two designs can be defined as:
$$[
B_{\mathrm{NoRegr}} :=
\big(
C(Y_{T_d}) = \text{True}
\Rightarrow
C(Y_{T_1\circ T_2}) = \text{True}
\big)
]$$

which is true whenever success of the direct design implies success of the decomposed one for the same instance.
The marginal probability
$$[
P_N(B_{\mathrm{NoRegr}} = \text{True}) \ge 1 - \varepsilon
]$$
thus quantifies how frequently the two-step design performs at least as well as the one-step composition within the modeled domain—an “almost-sure” (minus $\varepsilon$) notion of design improvement~\cite{brabermannapoli}. High probability values indicate that regressions (cases where the direct approach succeeds but the composed one fails) are rare in the selected domain.

A complementary variable,
$$[
B_{\mathrm{imp}} :=
\big(
C(Y_{T_d}) = \text{False}
\ \wedge \
C(Y_{T_2\circ T_1}) = \text{True}
\big)
]$$
captures instances where the new design achieves coverage un\-at\-tained by the original one.
High values of $P_N(B_{\mathrm{imp}} = \text{True})$ therefore indicate measurable improvement in end-to-end capability, expressed within the same probabilistic representation that encodes behavioral prescriptions and causal relations among transformation components.

\section{Future Work}

\paragraph{Reasoning Backbone.}
The proposed language is primarily intended for \emph{communication and documentation} rather than automated inference or verification. 
Nevertheless, its probabilistic foundations suggest natural paths toward analytical integration. 
Methods from probabilistic graphical modeling~\cite{DBLP:books/daglib/0023091} and probabilistic programming~\cite{probprogramming} could support empirical or symbolic reasoning once exogenous variables and transformation implementations are specified. 
For example, if inputs were generated through controlled sampling (e.g., from a benchmark suite), forward sampling could enable approximate inference over queries defined within a Generation Network.

\paragraph{Modeling Extensions.}
Future extensions will target the modeling of complex generation trajectories where the number of variables and interaction depth are not fixed a priori. 
We also plan to explore \emph{causal abstraction}~\cite{DBLP:conf/aaai/BeckersH19} and classical refinement~\cite{LG86} to enable hierarchical and modular representations, scaling Generation Networks from local patterns to full architectural designs.

\paragraph{Advanced Use Cases: Compositional Analysis}
A particularly promising direction concerns the use of Generation Networks to articulate verification desiderata involving distributional or representational compatibility between chained transformations.
In this view, one could analyze whether the output characteristics of a transformation align well with predefined input (sub)domain manifold representations under which a subsequent transformation has been extensively verified—an attempt to capture forms of \emph{relative compositionality}, non-regressive evolution of composed systems and hierarchical verification. Ideas from DLL testing based on latent spaces like~~\cite{DBLP:conf/icse/KangFY20, mozumder2025rbt4dnnrequirementsbasedtestingneural}\footnote{Note that variables denoting elements of latent spaces, encoding and decoding transformations as required for a VAE-like ID detection or generation can be already naturally modeled in GN.} in the setting of OOD for text (e.g.~\cite{wu-etal-2023-multi}) could be inspiring to define such mechanisms thus enable documenting and reasoning about how training, deployment, and interconnection assumptions affect overall system robustness and compatibility.

\section{Conclusions}
This work 
introduces Generation Networks (GNs), a graphical probabilistic language for documenting and reasoning about LLM-native software architectures.  
Rather than defining a new inference mechanism, the contribution lies in a structured notation that captures how algorithmic and LLM-based transformations interact within generative workflows. 
By integrating probabilistic and causal notions, GNs make it possible to express stochastic dependencies and design-level properties.  
The examples illustrate how higher-order relations among outcomes can be encoded, enabling reasoning about alternative designs, regressions, and performance gains within a unified probabilistic setting.

Looking ahead, GNs can serve as a foundation for design-level analysis and tool support in the emerging field of LLM-native software engineering.  
In this sense, the approach aspires to connect the abstraction principles of software engineering with the analytical rigor and toolsets of probabilistic graphical modeling—toward a principled, model-based engineering discipline for generative and learning-based systems.

\begin{acks}
Amazon Research Award -- Fall 2023 on Automated Reasoning; UBACyT 20020220300079BA and 20020190100126BA.
\end{acks}

 \bibliographystyle{ACM-Reference-Format}
 \bibliography{main}
\end{document}